\documentclass[12pt,preprint]{aastex}

\newcommand{\etal }{{et al.} }
\newcommand{\msun}{\thinspace M_\odot} 
\newcommand{\mjup}{\thinspace M_{\rm Jup}} 
\newcommand{\rsun}{\thinspace R_\odot} 
\newcommand{\vect}[1]{\mbox{\boldmath$#1$}}
\def\lesssim{\mathrel{\hbox{\rlap{\hbox{\lower4pt\hbox{$\sim$}}}\hbox{$<$}}}}
\def\gtrsim{\mathrel{\hbox{\rlap{\hbox{\lower4pt\hbox{$\sim$}}}\hbox{$>$}}}}
\newcommand{\cm}{\,{\rm cm}^{-3} } 
\newcommand{\km}{\,{\rm km\, s}^{-1}} 
\newcommand{\nc}{n_{\rm c} } 
\newcommand{\rcri}{R_{\rm c} }
\newcommand{\mdot}{M_\odot\,{\rm yr}^{-1} }
\newcommand{\tc}{t_{\rm c}}

\newcommand{\dfrac}[2]{{\displaystyle \frac{#1}{#2}} }

\shorttitle{Circumstellar Disk Formation}
\shortauthors{Machida \etal 2009}

\begin{document}
\title{Formation Process of the Circumstellar Disk: Long-term Simulations in the Main Accretion Phase of Star Formation}
\author{Masahiro N. Machida\altaffilmark{1}, Shu-ichiro Inutsuka\altaffilmark{2}, and Tomoaki Matsumoto\altaffilmark{3}} 
\altaffiltext{1}{National Astronomical Observatory of Japan, Mitaka, Tokyo 181-8588, Japan; masahiro.machida@nao.ac.jp}
\altaffiltext{2}{Department of Physics Nagoya University Furo-cho, Chikusa-ku Nagoya, Aichi 464-8602; inutsuka@nagoya-u.jp}
\altaffiltext{3}{Faculty of Humanity and Environment, Hosei University, Fujimi, Chiyoda-ku, Tokyo 102-8160, Japan; matsu@i.hosei.ac.jp}

\begin{abstract}
The formation and evolution of the circumstellar disk in unmagnetized molecular clouds is investigated using three-dimensional hydrodynamic simulations from the prestellar core until the end of the main accretion phase.
In collapsing cloud cores, the first (adiabatic) core with a size of $\sim10$\,AU forms prior to the formation of the protostar.
At its formation, the first core has a thick disk-like structure, and is mainly supported by the thermal pressure.
After the protostar formation, it decreases the thickness gradually, and becomes supported by the centrifugal force.
We found that the first core is a precursor of the circumstellar disk.
This indicates that the circumstellar disk is formed before the protostar formation with a size of $\sim10$\,AU, which means that no protoplanetary disk smaller than $<10$\,AU exists.
Reflecting the thermodynamics of the collapsing gas, at the protostar formation epoch, the first core (or the circumstellar disk) has a mass of $\sim 0.01-0.1 \msun$, while the protostar has a mass of $\sim 10^{-3}\msun$.
Thus, just after the protostar formation, the circumstellar disk is about $10-100$ times more massive than the protostar.
Even in the main accretion phase that lasts for $\sim10^5$\,yr, the circumstellar disk mass dominates the protostellar mass.
Such a massive disk is unstable to gravitational instability, and tends to show fragmentation.
Our calculations indicate that the planet or brown-dwarf mass object may form in the circumstellar disk in the main accretion phase.
In addition,  the mass accretion rate onto the protostar shows strong time variability that is caused by the perturbation of proto-planets and/or the spiral arms in the circumstellar disk.
Such variability provides a useful signature for detecting the planet-sized companion in the circumstellar disk around very young protostars.
\end{abstract}

\section{Introduction}
\label{sec:intro}
We believe that stars are born with a circumstellar disk.
The formation of the circumstellar disk is coupled with ``the angular momentum problem'' that is a serious problem in the star formation process, and  the dynamics of disks may determine the mass accretion rate onto the protostar that determines the final stellar mass.
In addition, planets are considered to form in the circumstellar (or protoplanetary) disk, and their formation process strongly depends on disk properties such as disk size and mass.
Thus, the formation and evolution of the circumstellar disk can provide a significant clue to star and planet formation.

Stars form in molecular clouds that have an angular momentum \citep{arquilla86,goodman93,caselli02}.
Thus, the appearance of a circumstellar disk is a natural consequence of the star formation process when the angular momentum is conserved in the collapsing cloud core.
In addition, observations have shown the existence of  circumstellar disks around the protostar \citep[e.g.,][]{watson07,dutrey07,meyer07}.
Numerous observations indicate that circumstellar disks around Class I and II protostars have a size of $\sim10-1000\,$AU and a mass of $\sim10^{-3}-0.1\msun$ \citep[e.g.,][]{calvet00,natta00}.
However, they correspond to phases long after their formation. 
Because the formation site of the circumstellar disk and protostar are embedded in a dense infalling envelope, we cannot directly observe newborn or very young circumstellar disks (and protostars).
Thus, we only observe the circumstellar disks long after their formation, i.e., around the class I or II protostar phase.

Observations also indicate that a younger protostar has a massive circumstellar disk \citep{natta00,meyer07}.
Recently, \citet{enoch09} observed a massive disk with $M_{\rm disk} \sim 1 \msun$ around class 0 sources, indicating that this massive disk can be present early in the main accretion phase.
However, unfortunately, observation cannot determine the real sizes of circumstellar disks, and how and when they form.
Therefore, we cannot understand the formation process of the circumstellar disk by observations.
The theoretical approach and numerical simulation are necessary to investigate the formation and evolution of the circumstellar disk.

Theoretically, the star formation process can be divided into two phases, i.e., the early collapse phase and main accretion phase.
The molecular cloud cores that are star cradles have a number density of $n \sim 10^4-10^6\cm$.
In the early collapse phase, the gas in the molecular cloud core continues to collapse until the protostar formation at $n \sim 10^{21}\cm$.
The protostar at its formation has a mass of $M_{\rm ps} \sim10^{-3}\msun$ that corresponds to the Jovian mass \citep{larson69,masunaga00}.
In the main accretion phase, the protostar acquires almost all its mass by the gas accretion to reach $\sim1\msun$.
In this paper, we define `the early collapse phase' as the period before the protostar formation, while `the main accretion phase' is defined as the period until the gas accretion onto the protostellar system (protostar and circumstellar disk) almost halts after the protostar formation.
In general, it is  considered that the circumstellar disk gradually increases its mass and size in the main accretion phase that is successively connected from the early collapse phase.
Thus, to understand the formation and early evolution of circumstellar disks, we should consider both the early collapse and main accretion phases; we need a self-consistent calculation from the collapse of the molecular cloud core until the end of the main accretion phase through the protostar formation.

However, to investigate the formation and evolution of circumstellar disks in numerical simulations,  we need a very long-term calculation with a sufficient spatial resolution, in which we should calculate the evolution of the protostellar system at least for the time comparable to the freefall timescale of the initial cloud core, i.e., $\gtrsim10^4-10^5$\,yr after the protostar formation.
In addition, we should resolve spatial scale length down to at least $\sim1$\,AU.
Reflecting the thermodynamics of the collapsing gas, two nested cores with a typical spatial scale appear in the early collapse phase \citep{larson69,masunaga00}. 
The inner core (so-called the second adiabatic core) corresponds to the protostar that has a size of $\sim1\rsun$, while the outer core that is called the first (adiabatic) core has a size of $\sim1-10$\,AU (for details, see \S\ref{sec:typical1}).
\citet{inutsuka09} expected that the circumstellar disk originates in the first core formed in the early collapse phase.
Thus, to investigate the formation of the circumstellar disk, we have to resolve the first core spatially; we need a spatial resolution of $\lesssim1$\,AU at least.
Moreover, we need a three dimensional calculation to properly treat the angular momentum transport in the circumstellar disk.

So far, many studies of star formation in the collapsing cloud mainly focused only on the early collapse phase \citep[e.g.,][]{bodenheimer00,goodwin07}.
To investigate the formation and evolution of the circumstellar disk, we have to calculate the early collapse and subsequent main accretion phases.
However, since such calculation requires a huge amount of CPU time, only a few studies reported the formation of the circumstellar disk in the collapsing cloud core including the main accretion phase.
\cite{kratter09} investigated the formation of the circumstellar disk with sub-AU resolution under the isothermal approximation, and showed frequent fragmentation of the disk.
\cite{walch09} also studied the circumstellar disk formation in the collapsing cloud core with a slightly coarser spatial resolution of $2$\,AU approximating the radiative cooling with adiabatic equation of state, and showed properties of the circumstellar disk.

In this study, using three-dimensional simulation adopting a barotropic equation of state with a higher spatial resolution than in previous studies, the evolution of the unmagnetized collapsing cloud cores is investigated until the gas accretion onto the protostar and circumstellar disk almost halts.
In three dimensions, we first calculate the evolution of the circumstellar disk by the end of the main accretion phase.
Although the circumstellar disk formation in the collapsing cloud may be investigated with a radiation-hydrodynamics code, it is very difficult to execute such a calculation even with current supercomputers because it takes a huge amount of CPU time.
Thus, to trace the gas thermodynamics, we chose a barotropic equation of state that is used even in recent two-dimensional simulations of the circumstellar disk formation \citep[e.g.,][]{vorobyov07,vorobyov09}.
The structure of the paper is as follows. 
The framework of our models and the numerical method are given in \S 2. 
The numerical results are presented in \S 3. 
We discuss the fragmentation condition of the circumstellar disk and implication of the planet formation in \S 4, and summarize our results in \S 5.

\section{Model Settings}
\label{sec:model}
To study the evolution of collapsing gas clouds and circumstellar disks, we solve the equations of hydrodynamics including self-gravity:
\begin{eqnarray} 
& \dfrac{\partial \rho}{\partial t}  + \nabla \cdot (\rho \vect{v}) = 0, & \\
& \rho \dfrac{\partial \vect{v}}{\partial t} 
    + \rho(\vect{v} \cdot \nabla)\vect{v} =
    - \nabla P -      \rho \nabla \phi, & 
\label{eq:eom} \\ 
& \nabla^2 \phi = 4 \pi G \rho, &
\end{eqnarray}
where $\rho$, $\vect{v}$, $P$, and $\phi$ denote the density, velocity, pressure, and gravitational potential, respectively. 
To mimic the temperature evolution calculated by \citet{masunaga00}, we adopt the piece-wise polytropic equation of state \citep[see,][]{machida07} as
\begin{equation} 
P =  c_{s,0}^2\, \rho \left[ 1+ \left(\dfrac{\rho}{\rho_c}\right)^{2/5} \right],
\end{equation}
%%%\begin{equation} 
%%%P = \left\{
%%%\begin{array}{ll}
%%% c_{s,0}^2 \rho & \rho < \rho_c, \\
%%% c_{s,0}^2 \rho_c \left( \dfrac{\rho}{\rho_c}\right)^{7/5} &\rho > \rho_c, 
%%%\label{eq:eos}
%%%\end{array}
%%%\right.  
%%%\end{equation}
 where $c_{s,0} = 190$\,m\,s$^{-1}$, and 
$ \rho_c = 3.84 \times 10^{-14} \, \rm{g} \, \cm$ ($n_c = 10^{10} \cm$). 

As the initial state, we take a spherical cloud with critical Bonnor--Ebert (BE) density profile, in which the uniform density is adopted outside the sphere ($r > \rcri$).
For the BE density profile, we adopt the central density of $\nc =  6 \times 10^{5}\cm$ and isothermal temperature of $T=10$\,K. 
For these parameters, the critical BE radius is $\rcri = 6.2\times10^3$\,AU.
To promote the contraction, we increase the density by a factor of $f$=1.68, where $f$ is the density enhancement factor that represents the stability of the initial cloud.
With $f=1.68$, the initial cloud has (negative) gravitational energy twice that of thermal energy.
Thus, the central density of the initial sphere is $n_{\rm c,ini} = 10^6\cm$, while the ambient density is $n_{\rm amb}=7.2\times 10^4\cm$.
We add $m=2$-mode non-axisymmetric density perturbation to the initial core.
Then, the density profile of the core is described as  
\begin{eqnarray}
\rho(r) = \left\{
\begin{array}{ll}
\rho_{\rm BE}(r) \, (1+\delta_\rho)\,f & \mbox{for} \; \; r < R_{c}, \\
\rho_{\rm BE}(R_c)\, (1+\delta_\rho)\,f & \mbox{for}\; \;  r \ge R_{c}, \\
\end{array}
\right. 
\end{eqnarray}
where $\rho_{\rm BE}(r)$ is the density distribution of the critical 
BE sphere, and $\delta_\rho$ is the axisymmetric density perturbation. 
For the $m=2$-mode, we chose
\begin{equation}
\delta_\rho = A_{\phi} (r/R_{\rm c})^2\, {\rm cos}\, 2\phi, 
\label{eq:dens-pert}
\end{equation}
where $A_{\phi}$ (=0.01) represents the amplitude of the perturbation.
The radial dependence is chosen so that the density perturbation remains regular at the origin ($r = 0$) at one time-step after the initial stage.
This perturbation ensures that the center of the gravity is always located at the origin.
The mass within $r < \rcri$ is $M = 1\msun$.
The gravitational force is ignored outside the host cloud ($r>\rcri$) to mimic a stationary interstellar medium.
Initially, the cloud rotates rigidly with angular velocity $\Omega_0$ around the $z$-axis.
We parameterized the ratio of the rotational to the gravitational energy ($\beta_0$) inside the initial cloud.
With different $\beta_0$, we calculated 10 models.
Model names, initial angular velocities $\Omega_0$, and $\beta_0$ are summarized in Table~1.

In the collapsing cloud core, we assume protostar formation occurs when the number density exceeds $n > 10^{13}\cm$ at the cloud center.
To model the protostar, we adopt a sink around the center of the computational domain.
In the region $r < r_{\rm sink} = 1\,$AU, gas having a number density of $n > 10^{13}\cm$ is removed from the computational domain and added to the protostar as a gravity in each timestep \citep[for details, see][]{machida09a}.
This treatment of the sink makes it possible to calculate the evolution of the collapsing cloud and circumstellar disk for a longer duration.

To calculate over a large spatial scale, the nested grid method is adopted \citep[for details, see][]{machida05a,machida06a}.
Each level of a rectangular grid has the same number of cells of $  128 \times 128 \times 16 $.
The calculation is first performed with five grid levels ($l=1$--$5$).
The box size of the coarsest grid $l=1$ is chosen to be $2^5 \rcri$.
Thus, a grid of $l=1$ has a box size of $1.97\times 10^5$\,AU.
A new finer grid is generated before the Jeans condition is violated.
The maximum level of grids is restricted to $l_{\rm max} \leqq 12$.
The $l=12$ grid has a box size of 96\,AU and cell width of 0.75\,AU.
With this method, we cover five orders of magnitude in spatial scale.

\section{Results}
\label{sec:results}
We investigated the cloud evolution and the circumstellar disk formation with different initial rotational energies, $\beta_0$.
Observations have shown that the molecular cloud cores have $10^{-4}\lesssim \beta_0 \lesssim 0.02$ with the mean value of $\beta_0 \sim 0.02$ \citep{goodman93,caselli02}.
In the following section, we show the cloud evolution and formation process of the circumstellar disk for two typical models (model 6 and 3).
Then, we compare the properties of circumstellar disks in clouds with different rotational energies.

\subsection{Typical Models}
\subsubsection{Cloud with $\beta_0=10^{-3}$}
\label{sec:typical1}
\begin{figure}
\includegraphics[width=150mm]{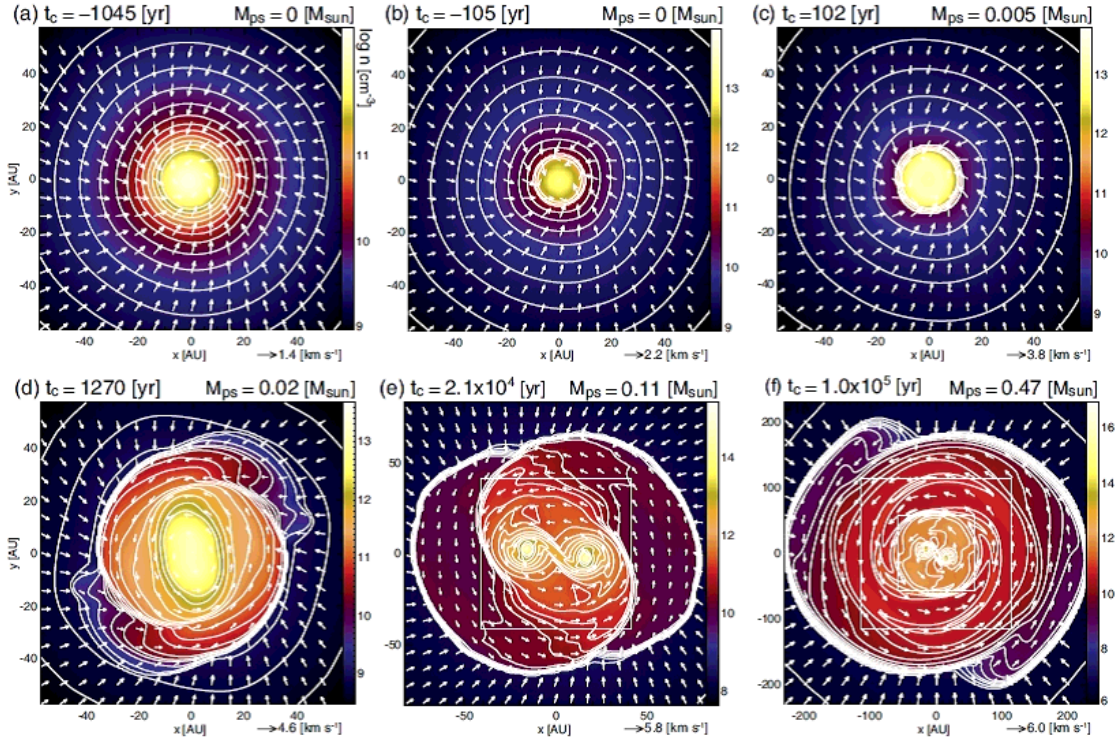}
\caption{
Evolution of the circumstellar disk for model 6 ($\beta_0 = 10^{-3}$).
The density ({\it color scale and contours}) and velocity ({\it arrows}) distributions on the equatorial ($z=0$) plane are plotted.
The spatial scale is different for each panel.
The time elapsed after the protostar formation ($t_{\rm c}$) and protostellar mass 
($M_{\rm ps}$) are shown in the upper part of each panel.
}
\label{fig:1}
\end{figure}

\begin{figure}
\includegraphics[width=150mm]{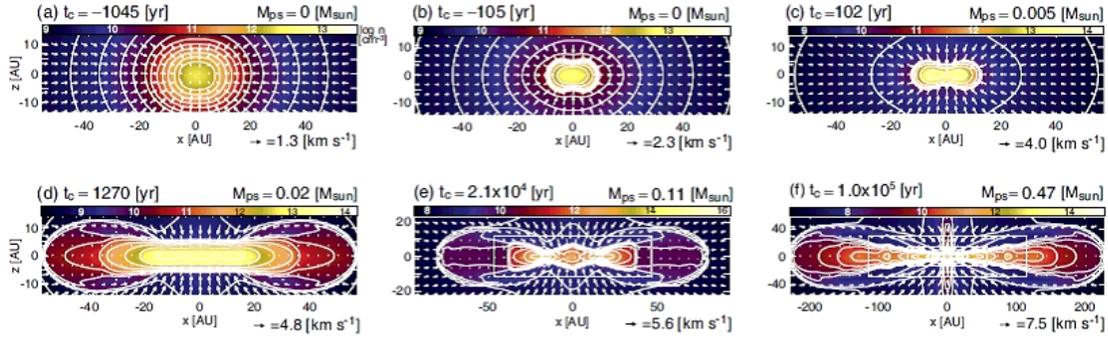}
\caption{
The density ({\it color scale and contours}) and velocity ({\it arrows}) distributions for the same model as in Fig.~\ref{fig:1} (model 6; $\beta_0=10^{-3}$) are plotted on the cross section of the $y=0$ plane at the same epochs as in Fig.~\ref{fig:1}.
}
\label{fig:2}
\end{figure}

Figures~\ref{fig:1} and \ref{fig:2} show the time sequence around the center of the cloud before (panels {\it a} and {\it b}) and after (panels {\it c} - {\it f}) the protostar formation  for the model with $\beta_0 =10^{-3}$ (model 6).
As denoted in \S\ref{sec:model}, we define the protostar formation epoch ($\tc = 0$) as the time when the maximum density reaches $n=10^{13}\cm$.
In these figures, the time after the protostar formation ($\tc$) and protostellar mass ($M_{\rm ps}$) are described in each panel.
The protostellar mass $M_{\rm ps}$ is derived as the mass falling into the sink.

\begin{figure}
\includegraphics[width=150mm]{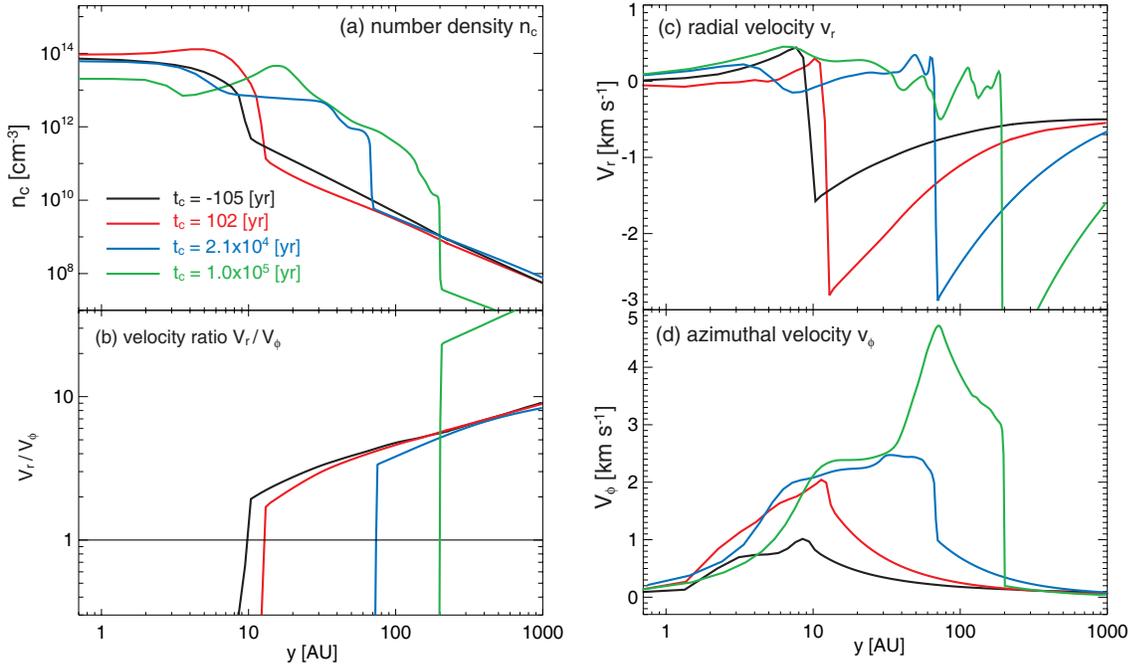}
\caption{
The radial distribution of ({\it a}) number density $n_{\rm c}$, ({\it b}) ratio of the radial to the azimuthal velocity $R_v$ (=$v_{\rm r}/v_{\phi}$), ({\it c}) radial  velocity $v_{\rm r}$, and ({\it d}) azimuthal velocity ($v_{\phi}$) at different epochs in model 6 ($\beta_0=10^{-3}$).
In panel {\it b}, the velocity ratio $R_v$ inside the disk is not displayed to stress the disk surface.
}
\label{fig:3}
\end{figure}

Panels {\it a}  and {\it b} in these figures indicate the formation of the disk-like structure before the protostar formation.
To clearly define the disk, we used the ratio of the radial to azimuthal velocity $R_v$ ($= \vert v_{\rm r}/v_{\phi} \vert$).
Figure~\ref{fig:3} shows the density distribution ({\it a}), the ratio of the radial to azimuthal velocity ({\it b}), and the distribution of the radial ({\it c}) and azimuthal ({\it d}) velocities against the distance from the protostar along the $y$-axis.
In Figure~\ref{fig:3}, these values for four different epochs corresponding to panels {\it b}, {\it c}, {\it e} and {\it f} of Figures~\ref{fig:1} and \ref{fig:2} are plotted.
Comparison of Figures~\ref{fig:1} and \ref{fig:2} with Figure~\ref{fig:3} {\it a} indicates that the shock front corresponds to the disk surface.
Thus, the disk-like structure is enclosed by the shock.
In Figure~\ref{fig:3}{\it a}, fine structures inside the shock surface at $t_c = 2.1 \times 10^4$\,yr and at $t_c = 1.0 \times 10^5$\,yr correspond to spiral density waves.
Figure~\ref{fig:3} ({\it c}) shows a sudden rise of the radial velocity at the shock front.
In addition, Figure~\ref{fig:3}{\it d} shows a large azimuthal velocity inside or near the shock front (i.e., inside or near the disk).
%%Note that the negative azimuthal velocity inside the disk (blue and green lines) is caused by the spin motion of fragment (see, Fig.~\ref{fig:1}).
Figures~\ref{fig:3}{\it c}  and {\it d} indicate that the gas rapidly falls into the center of the cloud with a slow rotation outside the disk, while the radial velocity slows and azimuthal velocity dominates inside the disk.
Therefore, the ratio of the radial to azimuthal velocity shows a sudden drop at the disk surface as shown in Figure~\ref{fig:3}{\it b}.
Note that, in Figure~\ref{fig:3}{\it b}, to stress the disk surface, the velocity ratio $R_v$ inside the disk is not displayed.
As a result, the velocity ratio $R_v$ is a good indicator to specify the disk.

In this paper, to determine the disk, we estimated the velocity ratio $R_v$ ($\equiv v_r/v_\phi$) in each cell, and specified the most distant cell having $R_v < 1$ from the center of the cloud.
Then, we defined the disk radius $r_{\rm disk}$ as the distance of the cell furthest from the origin, and the disk boundary density $\rho_{\rm d,b}$ as the density of the most distant cell.
Finally, we defined the disk that has a density of $\rho > \rho_{\rm d,b}$.
Thus, in our definition, the region inside the rapid drop of $R_v$ corresponds to the disk in Figure~\ref{fig:3}{\it b}.
Comparison of Figure~\ref{fig:3}{\it b} with Figures~\ref{fig:1}, \ref{fig:2} and \ref{fig:3}{\it a} indicates this definition of the disk well corresponds to a real size of the disk in simulation.

Figure~\ref{fig:2} shows that a thick disk-like structure formed before the protostar formation (Fig.~\ref{fig:2} {\it a} and {\it b}) transforms into a sufficiently thin disk after the protostar formation (Fig.~\ref{fig:2} {\it c} -- {\it f}).
As described in \S\ref{sec:intro}, theoretically, the star formation process can be divided into two phases: the early collapse phase (or the early phase of the star formation) and the main accretion phase (or the later phase of the star formation).
The early collapse phase was investigated in detail by \citet{larson69} and \citet{masunaga00}.
Here, we briefly describe this.
After the gas collapse is initiated in the molecular cloud core, the collapsing gas obeys the isothermal equation of state with temperature of $\sim 10$K for $n_c \lesssim 10^{10}\cm$ (isothermal phase).
Then cloud collapses adiabatically  ($ 10^{10} \lesssim n_c \lesssim 10^{16}$; adiabatic phase) and quasi-static core (i.e., first core) forms during the adiabatic phase.
After central density reaches $n_c \simeq 10^{16}\cm$, the equation of state becomes soft reflecting the dissociation of hydrogen molecules at $T \simeq 2\times 10^3$ K, and the gas collapses rapidly again, i.e., the second collapse begins. 
Finally, when the gas density reaches $n_c \simeq 10^{21} \cm$, the gas collapse stops and the protostar (or the second core) forms.
At this epoch, the early collapse phase ends and the main accretion phase begins.
In the main accretion phase, \citet{masunaga00} expected that the first core without the angular momentum disappears in $\sim100$\,yr after the protostar formation, while \citet{saigo06} pointed out that the first core having the angular momentum does not disappear in such short duration because the first core is supported by the rotation.
Thus, we can expect that the rotating first core formed in the early collapse phase becomes the circumstellar disk in the main accretion phase.
In the main accretion phase, the gas with the angular momentum continues to accrete onto the first core (or the circumstellar disk).

Figure~\ref{fig:2} clearly shows that the first core becomes the circumstellar disk in the main accretion phase; the first core is a precursor of the circumstellar disk.
The region enclosed by the shock in Figure~\ref{fig:1}{\it a} and Figure~\ref{fig:2}{\it a} corresponds to the first core that is formed after the gas becomes adiabatic.
Figures~\ref{fig:2}{\it b}-{\it e} show that the first core (or the circumstellar disk) gradually becomes thin with time.
Finally, a sufficiently thin disk appears as shown in Figure~\ref{fig:2}{\it f}.
Figure~\ref{fig:3}{\it b} shows that the disk has a radius of $\sim 10$\,AU before the protostar formation.
The disk extends up to $\sim 200$\,AU for $\tc \sim 10^5$\,yr in the main accretion phase.
In addition,  89\% of the total mass (i.e., 89\% of the initial host cloud mass) accretes onto the protostar and circumstellar disk system by this epoch ($\tc \lesssim 10^5$\,yr, see Table~1).

In summary, when we defined the circumstellar disk as the rotating disk around the protostar, the circumstellar disk with a size of $\sim10 - 20$\,AU  is already formed before the protostar formation.
This disk size at the protostar formation corresponds to that of the first core with rotation \citep{matsu03,saigo06}.
Therefore, the circumstellar disk has a minimum size of $\sim10$\,AU which implies that we cannot observe a disk with $\lesssim 10$\,AU even around very young protostars.

As shown in Figure~\ref{fig:1},  fragmentation occurs and two clumps form in the circumstellar disk $\sim 1.5\times 10^4$\,yr after the protostar formation.
At the fragmentation epoch, the protostellar mass is $M_{\rm ps}\sim 0.1\msun$, while the disk mass is $M_{\rm disk}\sim0.18\msun$.
Thus, the disk is 1.8 times more massive than the protostar.
As shown in \citet{larson69} and \citet{masunaga98}, at the protostar formation epoch, the protostar only has a mass of $\sim10^{-3}\msun$ that corresponds to Jovian mass.
This Jovian mass protostar acquires its mass by the gas accretion for $\sim10^5-10^6$\,yr to reach the solar mass object.
On the other hand, the first core has a mass of $\sim0.01-0.1\msun$ \citep{matsu03}.
The mass of the protostar and first core correspond to the Jeans mass at their formation epoch.
Figures~\ref{fig:1}--\ref{fig:3} show that the first core evolves directly into the circumstellar disk through the early  collapse to the main accretion phases.
Thus, during the early main accretion phase, the circumstellar disk is more massive than the protostar, as described in \citet{inutsuka09}.
Since such a massive disk is gravitationally unstable, fragmentation tends to occur \citep{toomre64}.
In this model, at the end of the calculation, each fragment is gravitationally bound, and has a mass $\sim0.037\msun$ with $\sim 12$\,AU of the separation between the central protostar and fragment.

\subsubsection{Cloud with $\beta_0=10^{-2}$}
\begin{figure}
%%%%%%%%%%
% Fig. 4 %
%%%%%%%%%%
\includegraphics[width=150mm]{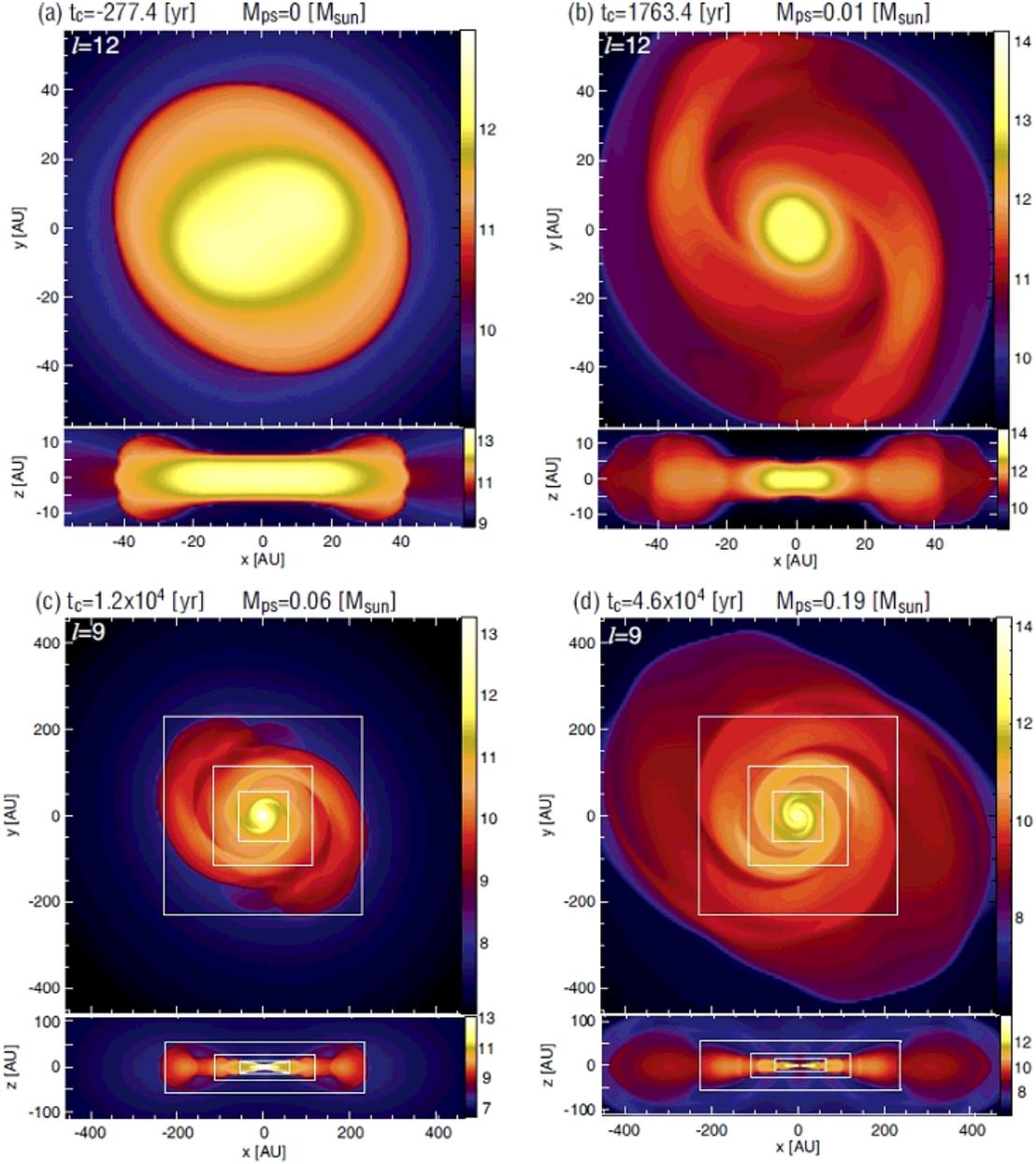}
\caption{
The density distributions ({\it color scale}) for model 3 ($\beta_0=10^{-2}$) are plotted on the cross section of the $z=0$ ({\it each upper panel}) and $y=0$ ({\it each lower panel}) planes.
The grid scale of lower panels is eight times as large as that of upper panels.
The time elapsed after the protostar formation ($t_{\rm c}$) and protostellar mass 
($M_{\rm ps}$) are shown in the upper side of each panel.
}
\label{fig:4}
\end{figure}

Figure~\ref{fig:4} shows the time sequence for the model with $\beta_0=10^{-2}$ (model 3).
Similar to the model with $\beta_0=10^{-3}$ (model 6), this model also shows the disk-like structure (i.e., disk-like first core) preceding the protostar formation (Fig.~\ref{fig:4}{\it a}).
Owing to the larger initial angular momentum, the first core in model 3 ($\beta_0 =10^{-2}$) has a flatter structure than that in model 6 ($\beta_0 =10^{-3}$).
For this model, at the protostar formation epoch, the protostar has a mass of $M_{\rm ps}\simeq10^{-3}\msun$, while the disk has a mass of $M_{\rm disk} \simeq0.1\msun$.
Thus, the disk is about 100 times more massive than the protostar.
Nevertheless, the circumstellar disk shows no fragmentation, as seen in Figure~\ref{fig:4}{\it b} - {\it d}.
Instead, the spiral structures appear with the gravitational instability in the circumstellar disk.
The circumstellar disk extends up to $\sim 500$\,AU (Fig.~\ref{fig:5}{\it d}) by the epoch at which the almost all gas in the host cloud has accreted onto the protostar and circumstellar disk system.
By this epoch,  the masses of protostar and circumstellar disk reach $M_{\rm ps} = 0.14\msun$ and $M_{\rm disk}=0.58\msun$, respectively.
Thus, even at end of the main accretion phase, the circumstellar disk is more massive than the protostar.
The disk properties and fragmentation condition are discussed in \S\ref{sec:dis1}.

\subsection{Disk Properties vs. Initial Rotational Energies}
\begin{figure}
\includegraphics[width=150mm]{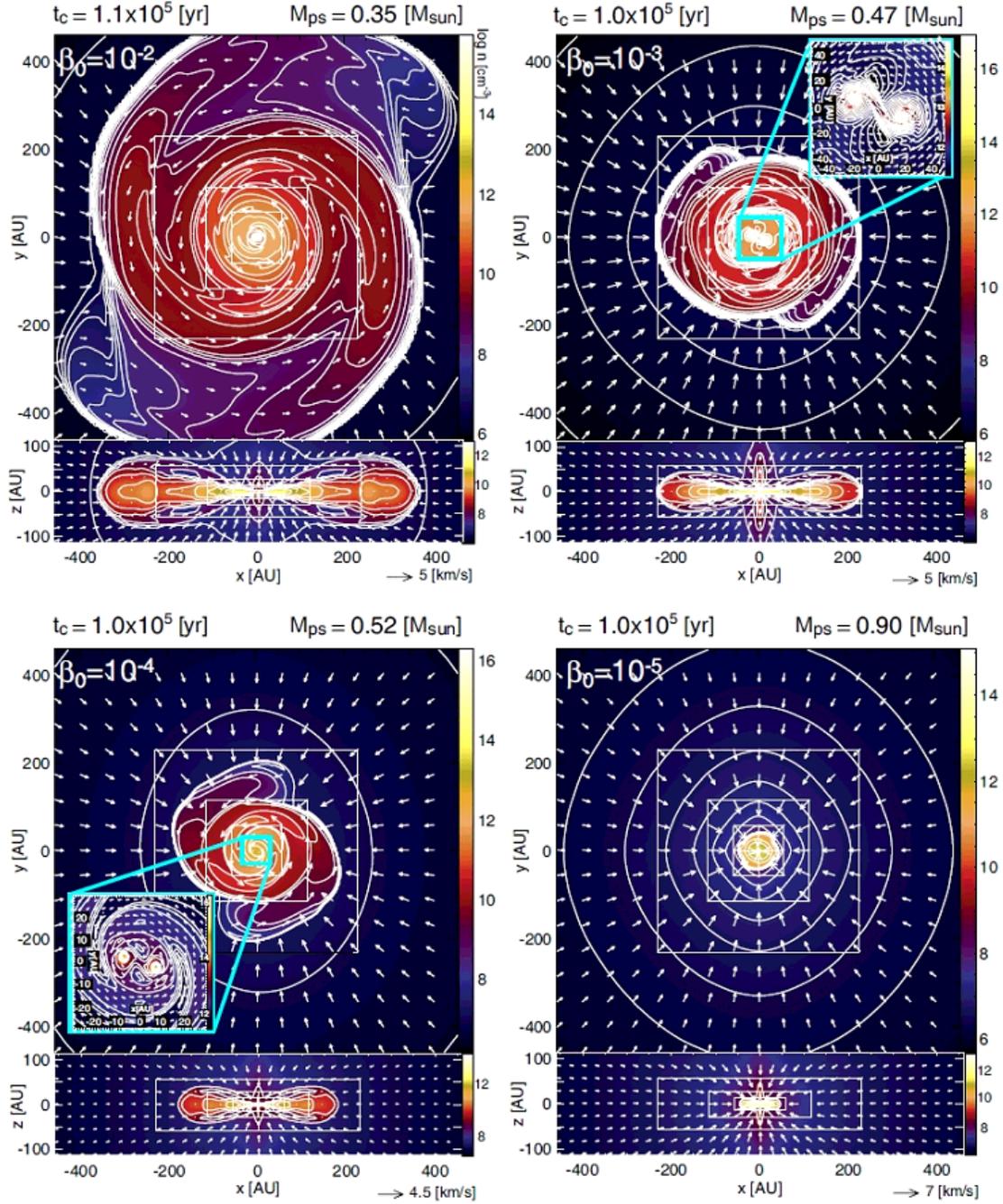}
\caption{
Final states on the cross section of $z=0$ for models with $\beta_0=10^{-2}$ (model 3), $10^{-3}$ (model 6), $10^{-4}$ (model 8), and  $10^{-5}$ (model 10).
The density ({\it color scale and contours}) and velocity ({\it arrows}) distributions on the equatorial ($z=0$) plane are plotted.
The time elapsed after the protostar formation ($t_{\rm c}$) and protostellar mass 
($M_{\rm ps}$) are shown in the upper side of each panel.
The close-up view of the central region is plotted for models showing disk fragmentation.
}
\label{fig:5}
\end{figure}

In this section, to investigate the disk properties, four different models with different initial rotational energies $\beta_0$ are presented.
Figure~\ref{fig:5} shows the density and velocity distribution for models with $\beta_0=10^{-2}$, $10^{-3}$, $10^{-4}$ and $10^{-5}$ at $\tc \simeq 10^5$\,yr after the protostar formation.  
Since the freefall timescale at the center of initial cloud is $t_{\rm ff,0} = 2.3\times 10^4$\,yr, they are structures at $\sim 4.3\,t_{\rm ff,0}$ after the protostar formation.
The figure shows a larger disk with larger $\beta_0$.
In addition, two clumps formed by fragmentation due to the gravitational instability appear in the circumstellar disk for models with $\beta_0=10^{-3}$ and $10^{-4}$, while no clump appears by the end of the main accretion phase for models with $\beta_0=10^{-2}$ and $10^{-5}$.
The fragmentation condition depends on the size and mass of the circumstellar disk (see, \S\ref{sec:dis1}).

\begin{figure}
%%%%%%%%%%
% Fig. 6 %
%%%%%%%%%%
\includegraphics[width=150mm]{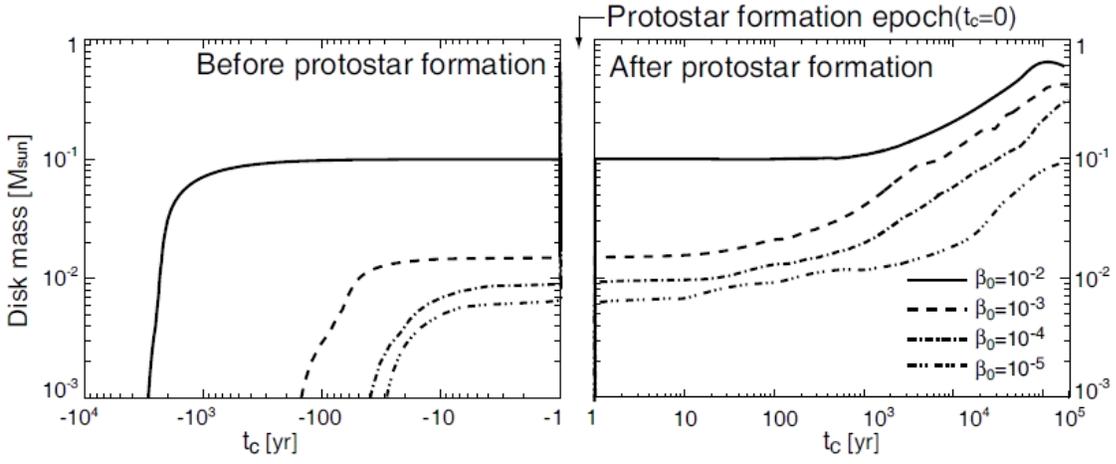}
\caption{
The evolution of the disk mass as a function of time before ({\it left}) and after ({\it right}) the protostar formation for models with $\beta_0=10^{-2}$ (model 3), $10^{-3}$ (model 6), $10^{-4}$ (model 8), and  $10^{-5}$ (model 10).
}
\label{fig:6}
\end{figure}

Figure~\ref{fig:6} shows the evolution of the disk mass before (left panel) and after (right panel) the protostar formation for the same models in Figure~\ref{fig:5}.
The figure indicates that the rotating disk forms before the protostar formation (i.e., $t_{\rm c} < 0$) and has a mass of $6\times10^{-3}-0.1\msun$ at the protostar formation epoch.
After the protostar forms, the circumstellar disk gradually increases its mass by gas accretion, and reaches $M_{\rm disk}=0.1-0.6\msun$ by the end of the main accretion phase.
The mass accretion rate onto the circumstellar disk during the main accretion phase ($\sim10^5$\,yr) is  $\dot{M}_{\rm disk}= (1  - 5)  \times 10^{-6}$\,$\msun$\,yr$^{-1}$.
This rate is determined by the balance between the mass accreting onto the circumstellar disk and mass infalling onto the protostar.

\begin{figure}
%%%%%%%%%%
% Fig. 7 %
%%%%%%%%%%
\includegraphics[width=150mm]{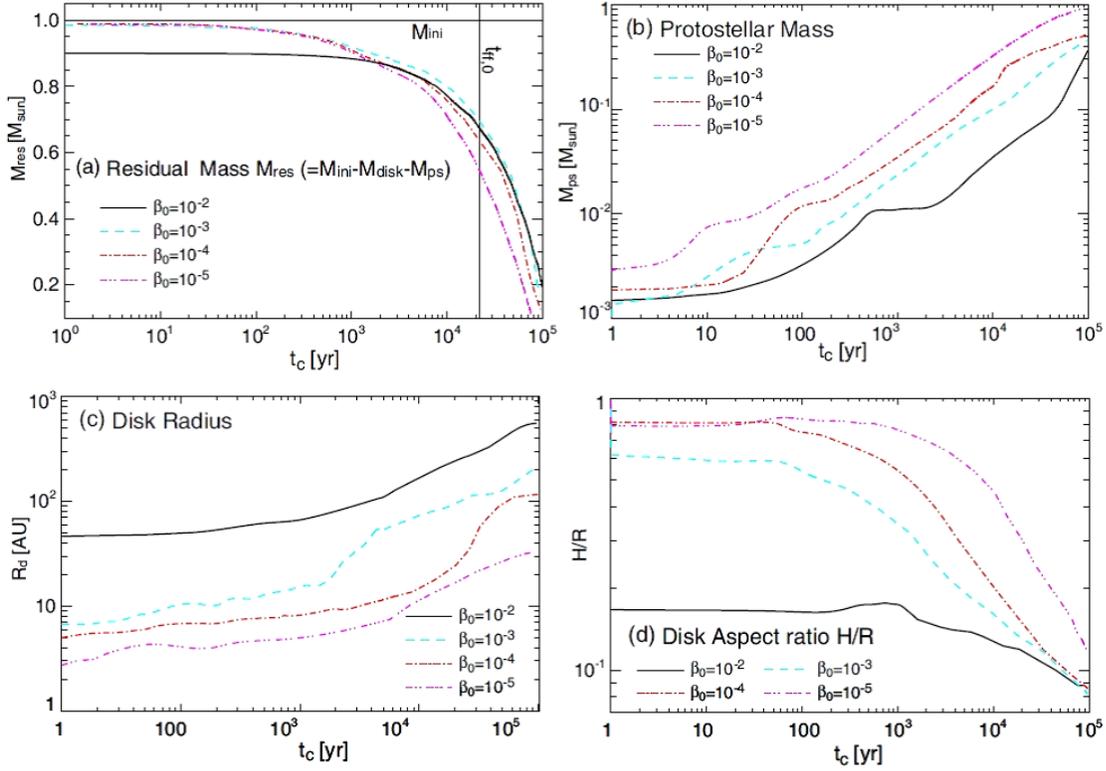}
\caption{
The evolution of ({\it a}) residual mass, $M_{\rm res}$ ($\equiv M_{\rm ini} - M_{\rm disk} - M_{\rm ps}$), ({\it b}) protostellar mass, $M_{\rm ps}$, ({\it c}) disk radius $R$, and ({\it d}) aspect ratio of the disk $H/R$ against time elapsed after the protostar formation for models with $\beta_0=10^{-2}$ (model 3), $10^{-3}$ (model 6), $10^{-4}$ (model 8), and  $10^{-5}$ (model 10).
}
\label{fig:7}
\end{figure}

Figure~\ref{fig:7}{\it a} plots a time sequence of the residual mass 
$ 
M_{\rm res} = M_{\rm ini} - M_{\rm disk} - M_{\rm ps}
$
for the same models in Figure~\ref{fig:5}, where $M_{\rm ini}$ is the initial mass of the host cloud (i.e., the mass inside $r<R_{\rm c}$ in the initial cloud).
The figure shows that the residual mass rapidly decrease in a freefall timescale and reaches $\sim 0.1 \msun$ in $\sim10^5$\,yr.
Thus, about 90\% of total mass accretes onto the protostar and circumstellar disk system by this epoch (see also, Table~1).
Therefore, the gas accretion almost halts and the main accretion phase ends at $\tc \sim 10^5$\,yr.

Figure~\ref{fig:7}{\it b} shows the time evolution of the protostellar mass for the same models.
The protostar has a mass of $(1-3)\times10^{-3}\msun$ at its formation epoch.
Then, the protostar acquires its mass by gas accretion and reaches $0.3-0.9\msun$ by the end of the main accretion phase.
The mass accretion rate onto the protostar during the main accretion phase is $\dot{M}_{\rm ps} = (3-9) \times 10^{-6}\,\msun$\,yr$^{-1}$ on average.
The time variability of the mass accretion rate is shown in \S\ref{sec:mdot}.
Figures~\ref{fig:6} and \ref{fig:7}{\it b} indicate that the model with a larger rotational energy has a larger $\dot{M}_{\rm disk}$, but smaller $\dot{M}_{\rm ps}$.
As shown in Figure~\ref{fig:5}, the cloud with a larger $\beta_0$ has a larger (or massive) rotating disk.
In such a disk,  the mass accretion onto the protostar is suppressed owing to larger angular momentum, and a less massive protostar appears with a relatively smaller protostellar accretion rate.

\begin{figure}
\includegraphics[width=150mm]{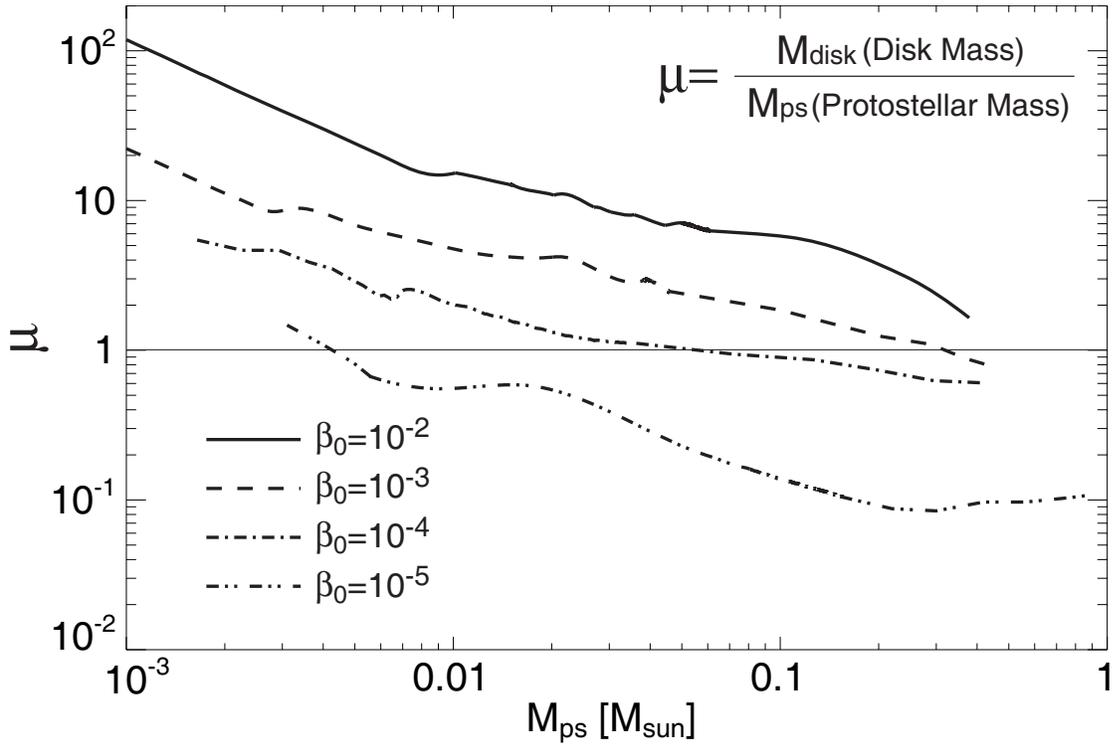}
\caption{
The evolution of the ratio of disk mass to protostellar mass against the protostellar mass for models with $\beta_0=10^{-2}$ (model 3), $10^{-3}$ (model 6), $10^{-4}$ (model 8), and  $10^{-5}$ (model 10).
}
\label{fig:8}
\end{figure}

Figure~\ref{fig:8} shows the disk-to-protostellar mass ratio $\mu$ against the protostellar mass for the same models in Figure~\ref{fig:5}.
The figure indicates that the circumstellar disk mass exceeds protostellar mass at the protostar formation epoch even for the model with a considerable small rotational energy.
The cloud with larger rotational energy has a relatively massive circumstellar disk at the protostar formation epoch.
For example, the circumstellar disk for model with $\beta_0  = 10^{-2}$ is about 100 times more massive than the protostar ($\mu=100$), while for the model with $\beta_0=10^{-5}$ it is twice as massive ($\mu=2$).
The circumstellar disk is more massive than the protostar for most of the main accretion phase, exception for the model with $\beta_0=10^{-5}$.
In addition, the circumstellar disk mass is comparable to the protostellar mass even at the end of the main accretion phase for models with $\beta_0 \ge 10^{-4}$.
Moreover, even in a cloud with $\beta_0=10^{-5}$ whose value is much smaller than the average of the observation $\beta_0=0.02$ \citep{caselli02}, the circumstellar disk has a 1/10 ($\mu=0.1$) of the protostellar mass at the end of the main accretion phase.

Figure~\ref{fig:7}{\it c} plots the evolution of the circumstellar disk radius after the protostar formation.
The figure shows that the circumstellar disk with the radius of $3-40$\,AU already exists at the protostar formation epoch (i.e., $\tc=0$).
The size of the circumstellar disk increases with time, and reaches $30-500$\,AU by the end of the main accretion phase.
The figure also indicates that the protostar has a larger circumstellar disk for the cloud with larger $\beta_0$.

Figure~\ref{fig:7}{\it d} plots the aspect ratio of the circumstellar disk, $H/R$, that is defined as
$ H/R \, \equiv \, h_z / (h _l h _s) ^{1/2} $, 
where $ h _l $, $ h _s $, and $ h_z$ are, respectively, lengths of the major, minor, and $z$-axes derived from the moment of inertia inside the circumstellar disk (see, \citealt{machida05a}).
Note that since $h_l$, $h_s$ and $h_z$ are weighted by the mass, they are slightly different from a real size of the circumstellar disk which is surrounded by the accretion shock as shown in Figures~\ref{fig:1} and \ref{fig:2}.
The figure shows that, for models with $\beta_0 < 10^{-2}$, the protostars have relatively thick circumstellar disks with $H/R>0.5$ at their formation epochs as also seen in Figure~\ref{fig:2}{\it c}.
On the other hand, for model with $\beta_0=10^{-2}$, the protostar at its formation already has a thin circumstellar disk with $H/R<0.2$.
However, in any model, the aspect ratio decreases with time and reaches $H/R<0.1$ by the end of the main accretion phase.
Thus, protostars have a sufficiently thin disk at the end of the main accretion phase.

\subsection{Mass Accretion Rate onto Protostar}
\label{sec:mdot}
\begin{figure}
\includegraphics[width=150mm]{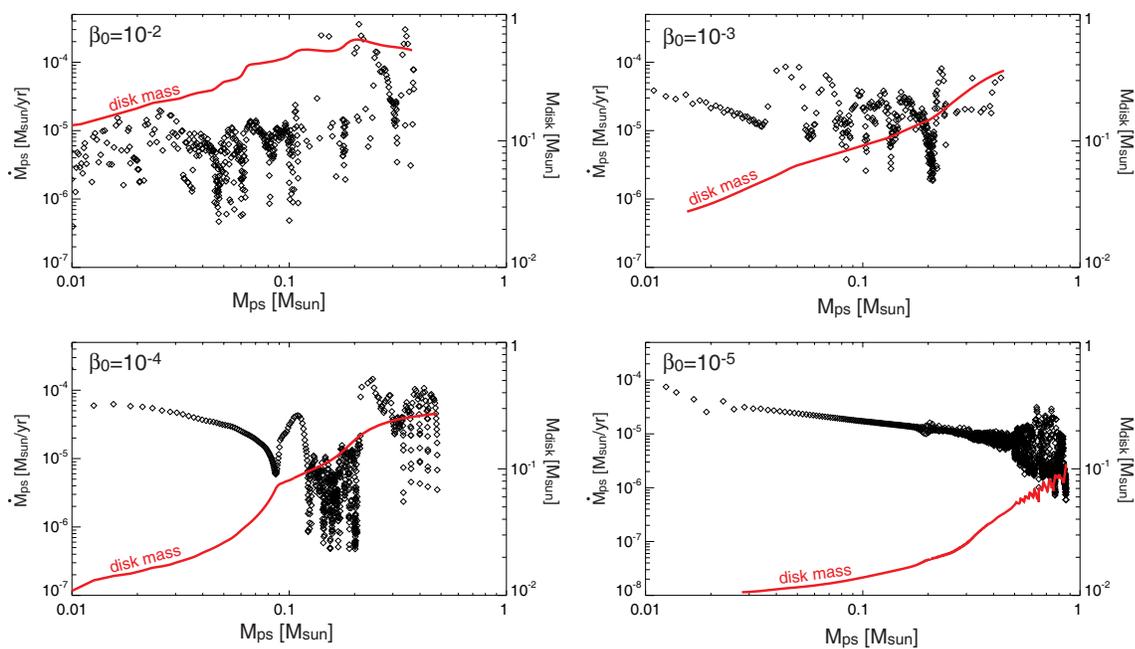}
\caption{
The mass accretion rate (left axis) and disk mass (right axis) are plotted for models 3, 6, 8 and 10. 
Note highly time-dependent $\dot{M}_{\rm ps}$ driven by gravitational torque due to non-axisymmetric structure in the disk.
}
\label{fig:9}
\end{figure}

Figure~\ref{fig:9} shows the mass accretion rate onto the protostar $\dot{M}_{\rm ps}$ (left axis) and circumstellar disk mass (right axis) against the protostellar mass.
As described in \S\ref{sec:model}, we removed the gas having a number density of $n > 10^{13}\cm$ in the region $r < 1\,$AU from the computational domain.
We regard the removed gas as the accreting mass onto the protostar, and estimated the mass accretion  rate in each step.
%%The mass accretion rate is defined as the mass falling into the region of $r<1$\,AU and 
The figure shows that the accretion rates are in the range of $10^{-4}  \lesssim  \dot{M}_{\rm ps} /(\mdot) \lesssim 10^{-6} $ in each model.
In the theoretical analysis of star formation, the mass accretion rate is prescribed as $\dot{M}=f\, c_s^3/G $, where $f$ is a numerical factor (e.g., $f=0.975$ for \citealt{shu77}, $f= 46.9$ for \citealt{hunter77}).
Since gas clouds have temperatures of $T=10$\,K ($c_s=0.2\km$), the accretion rate in the main accretion phase is $\dot{M}=(2-90)\times 10^{-6}\mdot$.
Thus, the accretion rate derived in our calculations corresponds well to the theoretical expectation.
However, the accretion rate shows strong time variability as the protostellar and circumstellar mass increases.

As seen in Figure~\ref{fig:9}, for models with $\beta_0 < 10^{-2}$, the accretion rate monotonically decreases initially, while it strongly fluctuates with time later as the protostellar (or circumstellar) mass increases.
In addition,  models with $\beta_0 = 10^{-2}$ show a strong time variability of the accretion rate right after the protostar formation.
This time variability stems from spiral arms or clumps caused by the gravitational instability in the circumstellar disk. 
The epoch at which the accretion rate begins to fluctuate corresponds to the formation epoch of the spiral arm or clump.
For example, for model with $\beta_0 =10^{-3}$, the accretion rate shows a slight time variability when  $M_{\rm ps} \lesssim 0.04\msun$ (see, Fig.~\ref{fig:1}{\it c} and {\it d}), while it shows a strong  variability when $M_{\rm ps} \gtrsim 0.04\msun$ (see, Fig.\ref{fig:1}{\it e}-{\it f}).
Indeed, when $M_{\rm ps} \sim 0.04\msun$, snapshot shows spiral structure.
Thus, this time variability of the mass accretion rate onto the protostar is closely related to the stability of the disk.
We discuss the disk stability in \S\ref{sec:dis1}.
%% and  will show periodicity of the mass accretion rate in a subsequent paper.

\section{Discussion}
\subsection{Stability and Fragmentation Condition for Circumstellar Disks}
\label{sec:dis1}
As shown in \S\ref{sec:results}, the circumstellar disk has a mass  larger than or comparable to the protostellar mass in the main accretion phase.
Such massive disk tends to show fragmentation or the formation of spiral structure that is closely related to the mass accretion onto the central star, subsequent evolution of the circumstellar and protoplanetary disk, and planet formation.
In this subsection, we discuss the disk stability and fragmentation condition.

As seen in Figure~\ref{fig:5}, for typical models (models 3, 6, 8, and 10), fragmentation occurs in clouds with moderate rotational energy of $\beta_0=10^{-3}$ and $10^{-4}$, while spiral arms appear without fragmentation in model with larger ($\beta_0=10^{-2}$) and smallest ($\beta_0=10^{-5}$) rotational energies.
To investigate the fragmentation condition of the disk,  we calculated Toomre's $Q$ parameter \citep{toomre64}, which is defined as 
\begin{equation}
Q = \dfrac{c_s \kappa}{\pi G \Sigma},
\end{equation}
where $c_s$, $\kappa$ and $\Sigma$ are the sound speed, the epicyclic frequency, and the surface density, respectively.
We also calculated the ratio of the critical Jeans length $\lambda_{\rm cri}$ to the disk radius $r_{\rm d}$ \citep{matsu03} as
\begin{equation}
R_{\rm c} = \dfrac{\lambda_{\rm cri}}{r_{\rm d}},
\end{equation}
where the critical Jeans length can be described as
\begin{equation}
\lambda_c = \dfrac{2 c_s^2}{G \Sigma} \left[ 
1 + (1-Q)^{1/2}
\right]^{-1}.
\end{equation}
It is considered that fragmentation occurs when both conditions of  $Q<1$ and $R_{\rm c}<1$ are fulfilled in the disk.
At each timestep, we estimated $Q$ and $\lambda_c$ in each cell inside the disk \citep[for detailed description, see][]{matsu03}.
Then, these values are averaged within the whole disk to roughly estimate the time variability of them.
%%stability and fragmentation condition for the circumstellar disk.
Note that although we may have to check local values of $Q$ and $R_{\rm c}$, not global (or averaged) values to investigate the disk stability and fragmentation \citep{kratter09}, we use global values to roughly estimate the time sequence of the disk stability.
%%We will investigate the disk properties (density and velocity distribution) and local stability (local $Q$) in a next paper.

\begin{figure}
%%%%%%%%%%
% Fig. 10 %
%%%%%%%%%%
\includegraphics[width=150mm]{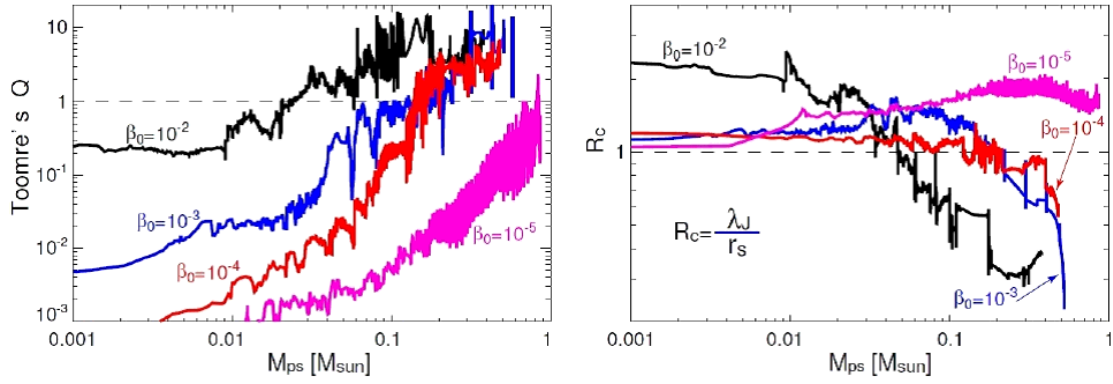}
\caption{
Averaged Toomre's Q (left panel) and the ratio of the critical Jeans length to disk radius (right panel) against the protostellar mass for models 3, 6, 8 and 10.
}
\label{fig:10}
\end{figure}

As shown in \S\ref{sec:results}, the circumstellar disk originates from the first core that is formed in the early collapse phase.
The first core is formed after the gas becomes adiabatic in the collapsing cloud core.
Thus, the first core is mainly supported by the thermal pressure gradient force at its formation.
Even when the initial host cloud has a significant rotational energy, the rotational energy of the first core is, at most,  only 10\% of the gravitational energy \citep{machida05a,machida06a}.
Then, the gas having a larger angular momentum accretes onto the disk (the first core or circumstellar disk) from the outer envelope as time goes on, and the disk gradually thins owing to the centrifugal force, and spreads to a large extent.
Thus, it is considered that $Q$ increases with time, because the gas with the larger specific angular momentum accretes onto the circumstellar disk with time; the epicyclic frequency or the ratio $\kappa/\Sigma$ is expected to become large with time. 
Figure~\ref{fig:10} left panel plots a time sequence of the averaged $Q$.
The figure shows that, for any model, $Q$ is much less than unity ($Q \ll 1$) at the protostar formation epoch and increases with time (or the protostellar mass).
Finally, it reaches order unity $Q\sim1$ by the end of the main accretion phase in any model.
The figure also shows that the circumstellar disk has a smaller $Q$ in the cloud with smaller rotational energy.
This is because, for the model with smaller $\beta_0$, the gas with smaller specific angular momentum forms a disk.

In general, fragmentation can occur in the rotating disk with $Q<1$.
However, for the model with $\beta_0=10^{-2}$, although the circumstellar disk has $Q<1$ when the protostellar mass is less than $M_{\rm ps}<0.02\msun$,  fragmentation does not occur, as shown in Figure~\ref{fig:4}.
This is because the disk radius is smaller than the critical Jeans length ($R_{\rm c}>1$) when $M_{\rm ps}<0.02\msun$.
On the other hand, for this model, the disk radius is larger than the critical Jeans length $R_{\rm c}<1$ when protostellar mass is larger than $M_{\rm ps}>0.4$,  while fragmentation does not occur because Q parameter exceeds unity $Q>1$ in this phase.
As a result, by the end of main accretion phase, this model shows no fragmentation, while the spiral structure appears as shown in Figure~\ref{fig:4}.

For the model with $\beta_0=10^{-3}$, the circumstellar disk keeps $Q<1$ for $M \lesssim 0.2\msun$, and the ratio $R_{\rm c}$ manages to reach $R_{\rm c}<1$ in this period.
Therefore, fragmentation occurs to form two clumps as shown in Figures~\ref{fig:1} and \ref{fig:2}.
Even in the model with $\beta_0 = 10^{-4}$, these two conditions ($Q<1$ and $R_{\rm c}<1$) are fulfilled for $M_{\rm ps} \lesssim 0.1\msun $, and fragmentation occurs.
On the other hand, the model with $\beta_0 = 10^{-5}$ has $Q<1$ through the main accretion phase, while fragmentation does not occur.
This is because the disk size is smaller than the critical Jeans length by the end of the main accretion phase.

\begin{figure}
%%%%%%%%%%
% Fig. 11 %
%%%%%%%%%%
\includegraphics[width=150mm]{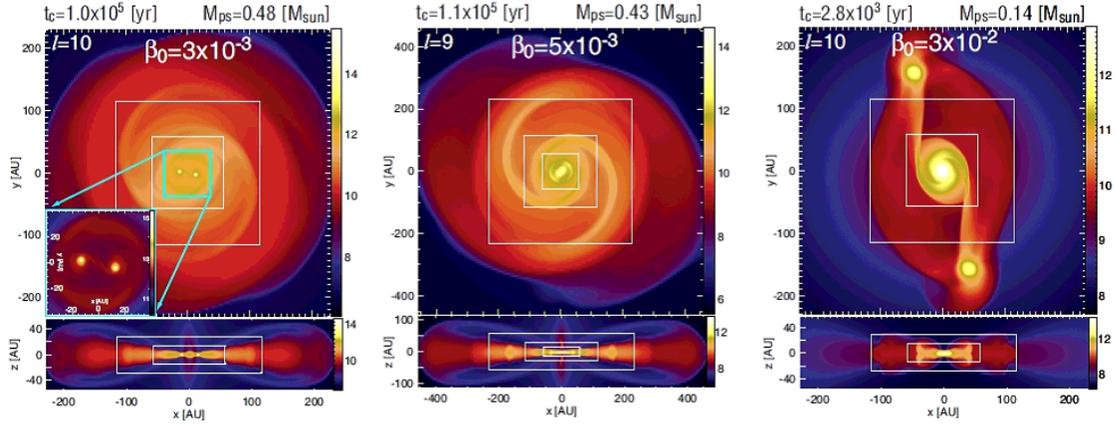}
\caption{
Final states on the cross section of $z=0$ (each top panel) and $y=0$ (each bottom panel) planes for models with $\beta_0 = 3 \times 10^{-3}$ (model 4), $ 5 \times 10^{-3}$ (model 3) and  $3 \times 10^{-2}$ (model 1).
The density distribution ({\it color scale}) is plotted in each panel.
The time elapsed after the protostar formation ($t_{\rm c}$) and protostellar mass 
($M_{\rm ps}$) are shown in the upper part of each panel.
}
\label{fig:11}
\end{figure}

To investigate the fragmentation condition in the main accretion phase, we calculated 10 models in total, as listed in Table~1.
Figure~\ref{fig:5} implies that the upper border of the rotational energy between fragmentation and non-fragmentation models exists in models between $\beta_0=10^{-2}$ and $10^{-3}$. 
To determine the fragmentation condition in more detail, we show the final fate for models $\beta_0=3\times10^{-3}$, $5\times10^{-3}$ and $3\times10^{-2}$ in Figure~\ref{fig:11}.
The figure indicates that fragmentation border exists in models between $\beta_0=3\times10^{-3}$ and $5\times10^{-3}$.
Fragmentation occurs in the model with $\beta_0=3\times 10^{-3}$, while fragmentation does not occur in the model with $\beta_0 = 5\times 10^{-3}$.
In addition, fragmentation occurs again in the model with $\beta_0=3\times 10^{-2}$.
Note that, for this model, we stopped calculation long before the end of main accretion phase ($\tc=2.8\times10^3$\,yr) because fragments move outwardly and the Jeans criterion is violated.
This model shows fragmentation in the early main accretion phase of $\tc=4\times10^3$\,yr (see, Table~1), because a sufficiently large disk is already formed in the early collapse phase (i.e., before the main accretion phase).
Moreover, the model with $\beta_0=5\times 10^{-2}$ shows fragmentation in the early collapse phase (i.e., before the protostar formation, see, Table~1).
The fragmentation process in the early collapse phase was reported in many past studies \citep[e.g.,][]{bodenheimer00, goodwin07}.
In summary, the fragmentation condition derived in our simulation can be described as follows:
\begin{enumerate}
\item $\beta_0 > 3\times 10^{-2}$ : fragmentation in the early collapse phase,
\item $10^{-5} < \beta_0 < 5\times 10^{-3}$ : fragmentation in the main accretion phase,
\item $\beta_0 < 10^{-5}$ and $5\times10^{-3} < \beta_0 < 3\times 10^{-2}$ : no fragmentation.
\end{enumerate}
Although models having parameters of $5\times10^{-3} < \beta_0 < 3\times 10^{-2}$ show no fragmentation, these models may show fragmentation in subsequent evolution stages.
As listed in Table~1, the mass of the circumstellar disk for no fragmentation models is comparable to the protostellar mass at the end of the main accretion phase, in which Toomre's $Q$ is nearly unity $Q\sim1$.
Such a disk can fragment to form several clumps when the cooling time scale is shorter than the Kepler rotational timescale in the protoplanetary disk \citep{gammie2001}.
In this study, we adopted a barotropic equation of state that makes the direct disk calculation from the molecular cloud until the end of the main accretion phase possible.
In the main accretion phase, the mass of the circumstellar disk overwhelms the protostellar mass with $Q\ll1$.
In such a situation, it is expected that the cooling process is not so important for fragmentation.
Instead, to investigate a marginal stable disk with $Q \sim1 $, the cooling process may be important \citep{durisen07}.
Ideally, we should investigate the circumstellar disk formation from the molecular cloud core with a radiative hydrodynamics, while we need to wait for further development of computational power to perform such calculations.
However, we can expect that fragmentation in the circumstellar disk generally occurs because a massive circumstellar disk inevitably appears in the star formation process.

\subsection{Implication for Planet Formation}
As described in \S\ref{sec:dis1}, a massive circumstellar disk inevitably appears in the main accretion phase.
This is because the first core that is a precursor of the circumstellar disk is about $10-100$ times more massive than the protostar, reflecting the thermodynamics of the collapsing gas.
Such a massive disk tends to show fragmentation owing to the gravitational instability \citep{durisen07}.
In our calculation, many models showed fragmentation and formation of small clumps in the circumstellar disk.
As listed in Table~1, at the end of the calculation, fragments have a mass of $M_{\rm frag} =  (19-72)\times10^{-3}\msun$ with a separation of $r_{\rm sep} = 7.1-94.2$\,AU.
For every model, fragments survived without falling into the central protostar.
Although the masses of fragments exist in the range of the brown dwarf ($M_{\rm frag}=13-75\,\mjup$), the planet-mass object may appear in the subsequent star formation phase, in which it is expected that the circumstellar disk cools to form less massive clumps \citep{durisen07}.
In this study, we  calculated the evolution of the circumstellar disk until the main accretion phase ends from the prestellar core phase.
However, further long-term calculations including an adequate radiative effect with a higher spatial resolution are necessary to properly investigate the mass and separation of fragments.
Since such calculation is beyond the scope of this study, we qualitatively, not quantitatively, discuss the planet formation in the subsequent evolution phase of the circumstellar disk below.

Recently, brown dwarf or planet mass companions have been observed.
\citet{neuhuser05} observed a brown-dwarf mass companion around GQ Lup with separation of $\sim100$\,AU.
In 2008, direct images of planet mass companions around Fomalhaut \citep{kalas08} and HR8799 \citep{marois08} are shown, in which they have masses of $3-10\mjup$ with separations of $24-98$\,AU.
It is considered that there are two different ways to form planets in the circumstellar (or protoplanetary) disk: one is the core accretion model  \citep{hayashi85}, the other is the gravitational instability model \citep{cameron78}.
In the core accretion scenario, it is difficult to account for the formation of massive planet in the region far from the protostar ($r_{\rm sep}\gtrsim5-10$\,AU) in the standard disk model \citep{perri74, mizuno78, hayashi85}.
On the other hand, a massive planet can form even in the region far from the protostar by fragmentation due to gravitational instability when the disk is sufficiently massive.
In this study, we showed that, in the circumstellar disk, gravitationally bound clumps can form in the region of $r_{\rm sep} > 5\,$AU.
In addition,  our calculation indicates that the massive circumstellar disks comparable to or more massive than the protostar form in the main accretion phase when the rotational energy of the molecular cloud is comparable to the observations ($10^{-4}< \beta_0 < 0.07$, \citealt{caselli02}).
Thus,  the gravitational instability scenario may be more common for the planet formation.

\subsection{Effects of Magnetic Field}
In this study, we ignored the magnetic field.
The magnetic field plays an important role in both the early collapse and main accretion phases, and  is related to the formation process of the star and circumstellar disk.
For example,  a large fraction of the cloud mass is blown away by the protostellar outflow that is driven by the Lorentz force from the circumstellar disk.
Thus, the protostellar outflow is expected to be closely related to the star formation efficiency \citep{matzner00,machida09a}.
The protostellar outflow is also related to the determination of the circumstellar disk mass \citep[e.g.,][]{inutsuka09}.
In addition, in the circumstellar disk, the turbulence owing to the magneto-rotational instability \citep[MRI;][]{balbus91} contributes to the angular momentum transfer that is related to the mass accretion rate onto the protostar.
Moreover, the magnetic field suppress fragmentation \citep{machida05b,machida08a,hennebelle08} and the formation of the disk \citep{mellon08,mellon09,duffin09,hennebelle09}.

However, even when the molecular cloud is strongly magnetized, the mass of the first core and protostar is, respectively,  $M_{\rm disk} \sim0.01-0.1\msun$ and $M_{\rm ps}\sim10^{-3}\msun$ \citep{machida07}.
Thus, as in the case of an unmagnetized cloud, the circumstellar disk mass overwhelms the protostellar mass at the protostar formation epoch.
This is because the mass and size of each object (first core and protostar) are determined by the thermodynamics of the collapsing gas \citep{masunaga00}.
In the main accretion phase, however, the masses of the circumstellar disk and protostar in magnetized clouds may be different from those in unmagnetized clouds because the angular momentum is transferred by the magnetic effect such as magnetic braking, protostellar outflow and MRI.
We will investigate the circumstellar disk formation in a magnetized cloud in detail in a subsequent paper.

\section{Summary}
To investigate the circumstellar disk formation and its properties, we calculated the evolution of the unmagnetized molecular clouds from the prestellar core phase until the end of the main accretion phase going through the protostar formation, using three-dimensional nested-grid simulation that covers both whole region of the initial molecular cloud and sub-AU structure.
We constructed 10 models with a parameter of the initial cloud's rotational energy ($\beta_0$), in which the size ($6.2\times10^3$\,AU) and mass ($1\msun$) of the initial molecular cloud are fixed, and calculated their evolution until about 90\% of the host cloud mass accretes onto either protostar or circumstellar disk (i.e., until the end of the main accretion phase).
The following results are obtained:

\begin{itemize}
\item 
{\it The first core is a precursor of the circumstellar disk.}
In the collapsing cloud core, the first core forms at $\nc \sim 10^{11}\cm$ with a size of $\sim10$\,AU before the protostar formation.
At its formation, the first core exhibits a thick disk-like configuration with aspect ratio of $H/R\simeq0.6-0.9$.
Then, after the protostar formation, the first core gradually transforms into a considerably thin disk ($H/R<0.1$) that corresponds to the circumstellar disk.
Thus, the minimum size of the circumstellar disk (or the size of the youngest circumstellar disk) is $\sim10$\,AU that corresponds to the size of the first core at its formation.
The circumstellar disk is mainly supported by the thermal pressure gradient force in the early collapse and early main accretion phases, and then it is mainly supported by the centrifugal force and becomes very thin in the later main accretion phase because the gas with a large angular momentum accretes onto it.
\item 
{\it The circumstellar disks are massive than the protostars throughout most of the main accretion phase.}
The first core has a mass of $\sim0.01-0.1\msun$, while the protostar has a mass of $\sim10^{-3}\msun$ at its formation.
The mass of each object almost corresponds to the Jeans mass at its formation.
The first core evolves directly into the circumstellar disk.
%% after the protostar formation.
Thus, at the protostar formation epoch, the circumstellar disk is inevitably more massive than the protostar: the circumstellar disk is about $10-100$ times more massive than the protostar.
After the protostar formation (i.e., in the mass accretion phase), the gas accretes onto the protostar through the circumstellar disk.
Since a part of the accreting gas accumulates in the  disk, not only the protostellar mass but also the circumstellar mass increases.
The mass growth rate of the protostar is larger than that of the circumstellar disk in the main accretion phase.
Nevertheless, the circumstellar disk is more massive than the protostar throughout the main accretion phase, reflecting the large initial mass ratio between the protostar and circumstellar disk at their formation.
\item 
{\it The planet and brown-dwarf mass objects tend to appear owing to the gravitational instability of the circumstellar disk in the main accretion phase.}
Since the circumstellar disk is massive than the protostar throughout the main accretion phase, the gravitational instability inevitably occurs in such massive disk.
In the cloud with initially moderate rotational energy ($10^{-5}< \beta_0 < 5\times10^{-3}$), fragmentation occurs in the circumstellar disk to form several clumps that have a mass of $20-70\mjup$ with a separation of $7-100$\,AU from the protostar.
On the other hand, in the cloud with initially larger rotational energy ($5\times10^{-3}< \beta_0 < 3\times10^{-2}$), fragmentation does not occur, but  spiral patterns appear in the main accretion phase.
Whether fragmentation or spiral arms appear depends on growth of the disk: fragmentation can occur when a sufficiently larger disk ($r_{\rm disk} > \lambda_{\rm cri}$) has sufficient mass ($Q<1$).
However, because the circumstellar disk not showing fragmentation in the main accretion phase has a mass comparable to the protostar, fragmentation may occur in later phase of the disk evolution, in which the disk cools and becomes more unstable again.
\item 
{\it The mass accretion rate onto the protostar shows strong time variability in the main accretion phase.}
After low-mass (planet or brown-dwarf mass) objects appear in the circumstellar disk, the mass accretion rate onto the protostar shows a strong time variability, because low-mass objects effectively transfer the angular momentum outwardly to promote gas accretion onto the protostar. 
On the other hand, there is little time variability of the accretion rate in a cloud with considerably lower rotation energy ($\beta_0=10^{-5}$) in the initial state.
In such a model, since neither clumps nor clear spiral patterns appear in the circumstellar disk, the angular momentum transport is not so effective in the present modeling without magnetic field.
Thus, the variability of the accretion rate provides a useful signature for detecting the low-mass companions or exo-planets around very young protostars.
\end{itemize}

Our results imply that, in the star formation process, the planet and brown-dwarf mass objects frequently appear in the circumstellar disk, indicating that the gravitational instability scenario may be a major pass for the planet formation.
To refine the formation scenario for planets, we need to study the disk evolution after the main accretion phase, in which both the radiative processes and magnetic effects should be taken into account.

\acknowledgments
Numerical computations were carried out on NEC SX-9 at Center for Computational Astrophysics, CfCA, of National Astronomical Observatory of Japan, and NEC SX-8 at the Yukawa Institute Computer Facility.
This work was supported by the Grants-in-Aid from MEXT (20540238, 21740136).

%%%%%%%%%%%%%
%%% Table %%%
%%%%%%%%%%%%%
\begin{table}
\setlength{\tabcolsep}{3pt}
\caption{Model parameters and calculation results}
\label{table}
\footnotesize
\begin{center}
%%\scalebox{.5}{%
\begin{tabular}{c|cc|cccccccccc} \hline
{\footnotesize Model} & $\beta_0$ &  $\Omega_0$ {\scriptsize [s$^{-1}$]}& $M_{\rm ps}$ {\scriptsize [$\msun$]} 
& $M_{\rm disk}$ {\scriptsize [$\msun$]}   &$M_{\rm red}$$^a$ {\scriptsize [$\msun$]} &$\mu$$^b$  & $r_{\rm disk}$ {\tiny [AU]} & $t_{\rm frag}$$^c$ {\scriptsize [yr]}
& $M_{\rm frag}$ {\scriptsize [$\msun$]} & $R_{\rm sep}$ [{\scriptsize \rm AU}] \\ \hline
1     & $5\times10^{-2}$ & $2.7 \times 10^{-13}$ & --- & ---   & --- & --- & --- & --- & 0.072 & 57.3\\

2     & $3\times10^{-2}$ & $2.1 \times 10^{-13}$ & 0.14 & 0.09   & 0.77 & 0.64 & 210 & $4.0\times 10^3$ & 0.061 & 94.2\\
3     & $10^{-2}$        & $1.2 \times 10^{-13}$ & 0.35 & 0.58   & 0.07 & 1.66 & 520 & ---   & --- & --- \\
4     & $5\times10^{-3}$ & $8.6 \times 10^{-14}$ & 0.43 & 0.51   & 0.06 & 1.66 & 409 & ---   & --- & --- \\
5     & $3\times10^{-3}$ & $6.6 \times 10^{-14}$ & 0.48 & 0.49   & 0.03 & 1.02 & 270 & $3.4\times 10^4$ & 0.058 & 21.1 \\
6     & $10^{-3}$        & $3.8 \times 10^{-15}$ & 0.47 & 0.42   & 0.11 & 0.89 & 202 & $1.5\times10^4$ &0.037 & 11.7 \\
7     & $3\times10^{-4}$ & $2.1 \times 10^{-14}$ & 0.51 & 0.38   & 0.11 & 0.74 & 152 & $9.8\times10^3$ &0.031 & 7.2 \\
8     & $10^{-4}$        & $1.2 \times 10^{-14}$ & 0.52 & 0.34   & 0.14 & 0.65 & 117 & $7.7\times10^3$ & 0.049 & 7.8 \\

9     & $3 \times 10^{-5}$ & $6.6 \times 10^{-15}$ & 0.80 & 0.16   & 0.04 & 0.20 & 29 & $2.0\times10^4$ & 0.019 & 7.1 \\
%%9     & $3 \times 10^{-5}$ & $6.6 \times 10^{-15}$ & 0.88 & 0.10   & 0.02 & 0.11 & 26 & $1.7\times10^4$ & 0.016 & 6.8 \\

10     & $10^{-5}$        & $3.8 \times 10^{-15}$ & 0.90 & 0.08   & 0.02 & 0.10 & 23  & --- & --- & ---\\
\hline
\end{tabular}
\end{center}
$^b$ Residual mass  [$=$ ($M_{\rm ini}$ $-$ $M_{\rm ps}$ $-$ $M_{\rm disk}$)]\\
$^a$ Disk-to-protostar mass ratio ($=$ $M_{\rm disk}$/$M_{\rm ps}$)\\
$^c$ Fragmentation epoch after the protostar formation
\end{table}

\clearpage
%%%%%%%%%%
% Fig. 1 %
%%%%%%%%%%

\clearpage
%%%%%%%%%%
% Fig. 2 %
%%%%%%%%%%

\clearpage
%%%%%%%%%%
% Fig. 3 %
%%%%%%%%%%

\clearpage

\clearpage
%%%%%%%%%%
% Fig. 5 %
%%%%%%%%%%

\clearpage

\clearpage

\clearpage
%%%%%%%%%%
% Fig. 8 %
%%%%%%%%%%

\clearpage
%%%%%%%%%%
% Fig. 9 %
%%%%%%%%%%

\clearpage

\clearpage

\end{document}